\pdfoutput=1

\documentclass[letterpaper,twocolumn,10pt]{article}

\usepackage{usenix2019,epsfig,endnotes}
\usepackage[hyphens]{url}
\usepackage[sort&compress,comma,square,numbers]{natbib}

\usepackage{color} 

\newcommand*\dash{\unskip\kern.16667em---\penalty\exhyphenpenalty
        \hskip.16667em\relax
}

 {\makeatletter
  \gdef\xxxmark{%
    \expandafter\ifx\csname @mpargs\endcsname\relax 
      \expandafter\ifx\csname @captype\endcsname\relax 
        \marginpar{\textcolor{red}{xxx}}
      \else
        \textcolor{red}{xxx~}
      \fi
    \else
      \textcolor{red}{xxx~}
    \fi}
  \gdef\xxx{\@ifnextchar[\xxx@lab\xxx@nolab}
  \long\gdef\xxx@lab[#1]#2{{\bfseries [\xxxmark \textcolor{red}{#2}
  ---{\scshape #1}]}}
  \long\gdef\xxx@nolab#1{{\bfseries [\xxxmark \textcolor{red}{#1}]}}
 }
 
 {\makeatletter
  \gdef\yyymark{%
    \expandafter\ifx\csname @mpargs\endcsname\relax 
      \expandafter\ifx\csname @captype\endcsname\relax 
        \marginpar{\textcolor{blue}{yyy}}
      \else
        \textcolor{red}{yyy~}
      \fi
    \else
      \textcolor{red}{yyy~}
    \fi}
  \gdef\yyy{\@ifnextchar[\yyy@lab\yyy@nolab}
  \long\gdef\yyy@lab[#1]#2{{\bfseries [\yyymark \textcolor{blue}{#2}
  ---{\scshape #1}]}}
  \long\gdef\yyy@nolab#1{{\bfseries [\yyymark \textcolor{blue}{#1}]}}
 }

\renewcommand{\paragraph}[1]{\vspace{0.1in}\noindent\textbf{#1.}}

\makeatletter
\def\@listi{\leftmargin\leftmargini
    \parsep 1\p@ \@plus0\p@ \@minus\p@
    \topsep 2\p@   \@plus0\p@ \@minus\p@
    \itemsep1\p@ \@plus0\p@ \@minus\p@}
\let\@listI\@listi\@listi
\makeatother

\usepackage{fancyhdr}
\begin{document}

\date{}

\title{\Large \bf Technology-Enabled Disinformation: Summary, Lessons, and Recommendations\\ \vspace{0.1in} \normalfont{\large\textit{December 2018}}}

\author{
{\rm John Akers} 
\and 
{\rm Gagan Bansal}
\and
{\rm Gabriel Cadamuro}
\and
{\rm Christine Chen}
\and 
{\rm Quanze Chen}
\and
{\rm Lucy Lin}
\and
{\rm Phoebe Mulcaire}
\and
{\rm Rajalakshmi Nandakumar}
\and 
{\rm Matthew Rockett}
\and
{\rm Lucy Simko}
\and
{\rm John Toman}
\and 
{\rm Tongshuang Wu}
\and
{\rm Eric Zeng}
\and
{\rm Bill Zorn} \vspace{0.001in}
\and 
{\rm Franziska Roesner} (franzi@cs.washington.edu)\vspace{0.1in} \\
Paul G. Allen School of Computer Science \& Engineering\\
University of Washington 
} 

\maketitle
\thispagestyle{fancy}


\begin{abstract}
    \vspace{0.05in}
    Technology is increasingly used \dash unintentionally (misinformation) or intentionally (disinformation) \dash to spread false information at scale, with potentially broad-reaching societal effects. For example, technology enables increasingly realistic false images and videos, and hyper-personal targeting means different people may see different versions of reality. This report is the culmination of a PhD-level special topics course\footnote{\url{https://courses.cs.washington.edu/courses/cse599b/18au/}} in Computer Science \& Engineering at the University of Washington's Paul G.~Allen School in the fall of 2018.
    The goals of this course were to study (1)~how technologies and today's technical platforms enable and support the creation and spread of such mis- and disinformation, as well as (2) how technical approaches could be used to mitigate these issues.
    In this report, we summarize the space of technology-enabled mis- and disinformation based on our investigations, and then surface our lessons and recommendations for technologists, researchers, platform designers, policymakers, and users.
    
\end{abstract}
\section{Introduction}

Misinformation is false information that is shared with no intention of harm. Disinformation, on the other hand, is false information that is shared with the intention of deceiving the consumer~\cite{councilOfEurope, LexiconOfLies}. This phenomenon, also referred to as information pollution~\cite{councilOfEurope} or false or fake news~\cite{Lazer2018TheSO}, is a critical and current issue. Though propaganda, rumors, misleading reporting, and similar issues are age-old, our current technological era allows these types of content to be created easily and realistically, and to spread with unprecedented speed and scale. The consequences are serious and far-reaching, including potential effects on the 2016 U.S. election~\cite{election}, distrust in vaccinations~\cite{vaccines}, murders in India~\cite{whatsapp-india}, and impacts on the lives of individuals~\cite{individual}. 

In this report, which is the culmination of a PhD-level special topics course\textsuperscript{1} in Computer Science \& Engineering at the University of Washington's Paul G.~Allen School in the fall of 2018, we consider the phenomenon of mis/disinformation from the perspective of computer science. The overall space is complicated and will require study from, and collaborations among, researchers from many different areas, including cognitive science, economics, neuroscience, political science, psychology, public health, and sociology, just to name a few. As technologists, we focused in the course, and thus in this report, on the technology. We ask(ed): what is the role of technology in the recent plague of mis- and disinformation? How have technologies and technical platforms enabled and supported the spread of this kind of content? And, critically, how can technical approaches be used to mitigate these issues \dash or how do they fall short?

In Section~\ref{sec:summary}, we summarize the state of technology in the mis/disinformation space, including reflecting on what is new and different from prior information pollution campaigns (Section~\ref{sec:overview}), surveying work that studies current mis/disinformation ecosystems (Section~\ref{sec:measure}), and summarizing the underlying enabling technologies (specifically, false video, targeting and tracking, and bots, in Section~\ref{sec:tech}), as well as existing efforts to curb the effectiveness or spread of mis/disinformation (Section~\ref{sec:defenses}). We then step back to summarize the lessons we learned from our exploration (Section~\ref{sec:lessons}) and then make recommendations particularly for technologists, researchers, and technical platform designers, as well as for policymakers and users (Section~\ref{sec:recommendations}).

Our overarching conclusions are two-fold. First, as technologists, we must be conscious of the potential negative impacts of the technologies we create, recognize that their designs are not neutral, and take responsibility for understanding and minimizing their potential harms. Second, however, the (causes and) solutions to mis/disinformation are not just based in technology. For example, we cannot just ``throw some machine learning at the problem''. In addition to designing technical defenses \dash or designing technology to attempt to avoid the problem in the first place \dash we must also work together with experts in other fields to make progress in a space where there do not seem to be easy answers. 
\section{Understanding Mis- and Disinformation}
\label{sec:summary}

A key goal of this report is to provide an overview of technology-enabled mis/disinformation\dash what is new, what is happening, how technology enables it, and how technology may be able to defend against it\dash which we do in this section. Others have also written related and valuable summaries that contributed to our understanding of the space and that we recommend the interested reader cross-reference~\cite{LexiconOfLies,MediaManipulation,councilOfEurope,Lazer2018TheSO}.

\subsection{What Is New?}
\label{sec:overview}

While propaganda and its ilk are not a new problem, it is becoming increasingly clear that the rise of technology has served to catalyze the creation, dissemination, and consumption of mis/disinformation at scale. Here we briefly detail, from a technology standpoint, the major factors that have contributed to our current situation:

\begin{enumerate}
\item \textbf{Democratization of content creation.} Anyone can start a website, blog, Facebook page, Twitter account, or similar and begin creating and sharing content. There has been a transformation from a highly centralized system with a few major content producers to a decentralized system with many content producers. This flood of information makes it difficult to distinguish between true and false information (i.e., information pollution).  Additionally, tools to create realistic fake content have become available to non-technical actors (e.g., DeepFakes~\cite{chesneyDeepFakes,deepfacelab}), significantly lowering the barrier to entry in terms of time and resources for people to create extremely realistic fake content.
\item \textbf{Rapid news cycle and economic incentives.} The more clicks a story gets, the more money gets made through ad revenue. Given the rapid news cycle and plethora of information sources, getting more clicks requires producing content that catches consumers' attention\dash for instance, by appealing to their emotions~\cite{affect-attitude} or playing into their other cognitive biases~\cite{cognitivebiases,biasCheatSheet}. In addition, social media platforms that provide free accounts to users make money via targeted advertising. This allows a variety of entities, with intentions ranging from helpful to harmful, to get their messages in front of users.
\item \textbf{Wide and immediate reach and interactivity.} Content created on one side of the planet can be viewed on the other side of the planet nearly instantaneously. Moreover, these content creators receive immediate feedback on how their campaigns are performing \dash and can easily iterate and A/B test \dash based on the number of likes, shares, clicks, comments, and other reactions on social media. Contrast the six long years that it took for the Russian disinformation campaign about AIDS having been created by the U.S. to take off in the 1980s ~\cite{operationInfektion,AIDS} with the viral spread of false stories like Pizzagate~\cite{pizzagate} during the 2016 U.S. election. 
\item \textbf{Organic and intentionally created filter bubbles.}
In today's web experience, individuals get to choose what content they do or do not want to see, e.g., by customizing their social media feeds or visiting specific websites for news. The resulting echo chambers are often called ``filter bubbles''~\cite{filterbubble} and can help mis/disinformation spread to receptive people who consume it within a limited broader context. Though some recent studies suggest that filter bubbles may not be as extreme as feared (e.g.,~\cite{filterbubble2}), others find evidence for increased polarization~\cite{polarization}. Moreover, cognitive biases like the backfire effect (discussed below) suggest that exposure to alternate views is not necessarily effective in helping people change their minds.
Worse, entities wishing to spread disinformation can intentionally \textit{create} filter bubbles by directly and accurately targeting (e.g., via ads or sponsored posts) potentially vulnerable individuals.
\item \textbf{Algorithmic curation and lack of transparency.} Content curation algorithms have become increasing complex and fine-tuned \dash e.g., the algorithm that determines which posts, in what order, are displayed in a user's Facebook news feed, or YouTube's recommender algorithms, which have been frequently criticized for pushing people towards more extreme content~\cite{youtube}. 
At the same time, these algorithms are not transparent for end users (e.g., Facebook's ad explanation feature~\cite{Andreou2018InvestigatingAT}).
This complexity can make it challenging for users to understand why they are having specific viewing experiences online, what the real source of information (e.g., sponsored content) is~\cite{Mathur2018EndorsementsOS}, and even \textit{that} they are having different experiences from other users. 


\item \textbf{Scale and anonymity in online accounts.}
Actors wishing to spread disinformation can leverage the weak identity and account management frameworks of many online platforms to create a large number of accounts (``sybils'' or bots) in order to, for example, pose as legitimate members of a political movement and/or artificially create the impression that particular content is popular (e.g.,~\cite{StarbirdBLM}). It is not clear how to address this issue, as stricter account verification can have negative consequences for some legitimate users~\cite{Haimson2016ConstructingAE}, and bot accounts can have legitimate purposes~\cite{goodbots,Ferrara2016TheRO}.

\end{enumerate}

\paragraph{What has not changed: Cognitive biases of human consumers}
Though the above technology trends have contributed to creating the currently thriving mis/disinformation ecosystem, ultimately, the content that is effective relies on something that has not changed: the cognitive biases of human consumers~\cite{cognitivebiases,biasCheatSheet}. A prescient article from 2002~\cite{Cybenko2002CognitiveHA} discussed how these cognitive biases can be exploited via technology, and such strategies are common in marketing~\cite{cognitivebiases} and traditional propaganda. A common such bias is ``confirmation bias'', by which people pay more attention and lend more credence to information that confirms something they are already inclined to believe. Another example that will recur in this report is the ``backfire effect'', which suggests that when confronted with facts intended to change a person's opinion or belief about something, this confrontation can have the opposite effect, causing people to more strongly hold their original belief~\cite{backfireEffect,backfireEffect2}.
A full survey of these exploitable human cognitive ``vulnerabilities'', though outside the scope of this report, is an important component of understanding why mis/disinformation campaigns are effective and how (not) to combat them.

\subsection{Measuring the Problem}
\label{sec:measure}

We begin by surveying work that has studied today's mis/ disinformation ecosystems directly, attempting to quantitatively and qualitatively understand the problem. Understanding the problem\dash who are the adversaries, who are the targets, what are the techniques used, and why are they effective\dash is crucial for ultimately developing effective defenses.

\subsubsection{Ecosystem Studies}

A number of academic studies have now examined various aspects of the mis/disinformation ecosystem. Many of these studies are case studies focusing on a particular issue, campaign, or time span (e.g.,~\cite{StarbirdEcosystem,StarbirdBLM,Fourney2017}). Some more general studies or surveys exist as well (e.g.,~\cite{Zubiaga2018,MediaManipulation,WhoSharedIt}). 
We summarize some of the key findings from this work here.

\paragraph{Techniques}
One repeated theme in this research is that sources of mis/disinformation rely on a range of techniques in terms of sophistication. 
For example, a study of disinformation campaigns embedded in Twitter conversations around Black Lives Matter suggested that social media accounts that spread mis/disinformation range from low-effort spambots, to slightly more human-seeming, semi-automated accounts with some degree of human supervision, to purely human-run accounts that build large followings before exploiting the trust of those social links to spread mis/disinformation widely~\cite{StarbirdBLM}.
Supporting these results more broadly, Marwick et al.~\cite{MediaManipulation} provide a partial survey of groups and movements that have been particularly effective in manipulating the media to spread their messages, using methods such as: impersonation of activists to exploit the trust aspect of social networks, bot promotion of a story on social media, and government-funded media outlets.

\paragraph{Structure}
Understanding how mis/disinformation content or information ecosystems differ from other information ecosystems can potentially help detect them. For example, a study of alternate news media coverage of the Syrian White Helmets found that these media often copy content without attribution; clusters of sources which shared articles reveal a densely-linked alternative media cluster which is largely disconnected from reputable news media~\cite{StarbirdEcosystem}.

\paragraph{Consumers}
 In addition to understanding disinformation campaigns, ecosystems studies can help us understand the behaviors of human consumers. For example, a study of the geographic and temporal trends of fake news consumption during the 2016 U.S. election found a correlation of aggregate voting behavior with consumption of fake news across states~\cite{Fourney2017} (though the causality is, of course, unclear).

More generally, evidence suggests~\cite{WhoSharedIt} that users base their trust in an article on their trust of the friend who they saw share it, i.e., information local to them, rather than the source or upstream links in the chain of shares. Further, people mistakenly believe that they take the trustworthiness of the source into account more than they actually do~\cite{MediaManipulation}. On some platforms, design choices mean that messages may spread in a way that obscures the source (e.g., WhatsApp) or conceals intermediate links in the chain between source and sharer (e.g., Twitter). These design choices may contribute to the spread of mis/disinformation and can make it difficult for external researchers to study the ecosystem.

\subsubsection{Tools for Measurement}


A major challenge in conducting measurement studies of mis/disinformation, or of attempting to counter it by fact-checking individual stories, is that it requires significant manual effort. To help with this process, several tools\dash for researchers, journalists, and/or end users\dash have been developed to surface the sources and trajectories of shared content or to facilitate fact checking directly. 

For example, Hoaxy~\cite{hoaxy} is a tool that collects and tracks mis/disinformation and corresponding fact checking efforts on Twitter. A study by its creators~\cite{Shao2016HoaxyAP} found that, perhaps unsurprisingly, fact checking content is shared significantly less than the original article. In another study~\cite{shao2}, its creators used Hoaxy to demonstrate the central role of social spambots in the mis/disinformation ecosystem\dash a valuable conclusion, as it suggests that curbing these bots may be crucial in the fight against mis/disinformation (see Section~\ref{sec:bots}).

TwitterTrails~\cite{TwitterTrails} is a related tool that provides information regarding the origin, burst characteristics, and propagation of content on Twitter. The resulting visualizations are intended to help users to investigate the veracity of the content. Hamilton 68~\cite{hamilton68} similarly provides a dashboard of information about Russian influence operations on Twitter.

\paragraph{Challenges}
Though valuable to helping us understand the broader ecosystem, none of these tools is yet a silver bullet. Some (e.g., TwitterTrails and Emergent~\cite{emergent}, a seemingly discontinued real-time rumor tracker) suffer from being insufficiently automated. Additionally, in some cases, it is not directly clear how the surfaced information should be used by researchers or end users, and worse, may be misunderstood~\cite{hamilton68-blog}.  Finally, these tools are all targeted at Twitter because of its public nature; closed platforms like Facebook do not support similar efforts.  These limitations are not due to the lack of effort or intention by the tools' creators but rather due to fundamental challenges. Mis/disinformation campaigns vary widely in their techniques, automatically detecting ``fake news'' is hard (as discussed below), humans suffer from fundamental cognitive biases, and while knowing the source and trajectory of content can help shed light on the dynamics of the ecosystem, measurement may not directly provide a solution. Nevertheless, such tools are a valuable component of a defender's toolbox.

\subsection{Enabling Technologies}
\label{sec:tech}

We now turn to several key enabling technologies that underlie the recent rise of mis/disinformation at scale: extremely realistic and easy-to-create false video (Section~\ref{sec:falsevideo}), individualized tracking of a user's browsing behaviors and hyper-personal targeting of content based on the results of this tracking (Section~\ref{sec:tracking}), and social media bots (Section~\ref{sec:bots}).

\subsubsection{False Video}
\label{sec:falsevideo}

False media has existed for as long as there has been media to falsify: forgers have faked documents or works of art, teenagers have faked driver's licenses, etc. With the advent of digital media, the problem has been amplified, with tools like Photoshop making it easy for relatively unskilled actors to perform sophisticated alterations to photographs. More recently, developments in AI have extended that capability to permit the creation of realistic false video.

The most popular technique to produce photorealistic false video is the reanimation of faces. In a typical setup, a target video, such as an announcement by a politician, is reanimated to show the same sequence of facial expressions as a source video, such as a political opponent sitting in front of a webcam in their living room. Usually the reanimation process is implemented with generative adversarial networks, and a significant amount of training data may be required. The resulting false video is often referred to as a ``deepfake''.

\paragraph{Instantiations}
Various implementations exist in the literature and publicly on the internet. The Face2Face tool, published in 2016~\cite{Thies2016Face2FaceRF}, is an early example of real-time facial reenactment. In 2017, the term ``deepfake'' went public on reddit, along with the FakeApp tool~\cite{deepfake,deepfacelab}. The Deep Video Portraits project significantly refined the process in 2018~\cite{Kim2018DeepVP}, requiring only a few minutes of video of the target to create a realistic real-time reanimation. Other work from 2017~\cite{Suwajanakorn2017SynthesizingOL} allows the reanimation to be automatically derived from an audio stream, without the need for a source video (but with the need for significant training video data of the target, e.g., Obama).

\paragraph{Implications}
Synthetic, photorealistic video is hardly new, but historically it has been prohibitively expensive to produce, limiting its appearance to scenarios like special effects in movies. Tools like FakeApp democratize this capability, making it possible for individuals to create similarly high-quality synthetic video in a restrictive but growing set of cases. And we can only expect synthetic video technology to improve, given its importance in the entertainment industry, from video games to Snapchat filters. In spite of its entertainment value, the increasing prevalence of synthetic video can have negative effects on our information ecosystem, from spreading mis/disinformation to undermining the credibility of legitimate information (allowing anyone to argue that anything has been faked). 

\subsubsection{Tracking and Targeting}
\label{sec:tracking}

The economic model of today's web is based on advertising: users receive content for ``free'' in exchange for seeing advertisements. To optimize these advertisements, they are increasingly targeted at specific consumers; to enable this targeting, widespread and sophisticated tracking of user browser behaviors has been developed. These technologies of tracking and targeted can be leveraged not only by legitimate traditional advertisers but also by entities wishing to spread mis/disinformation for monetary gain (e.g., viral content to maximize clicks) as well as directly malicious purposes (e.g., affecting the outcome of an election).


\paragraph{Tracking}
There has been significant prior work studying tracking and targeting within the computer security and privacy community, which we very briefly summarize here. At the highest level, so-called third party tracking is the practice by which companies embed content \dash like advertising networks, social media widgets, and website analytics scripts \dash in the first party sites that users visit directly. This embedding relationship allows the third party to re-identify users across different websites and build a browsing history. In addition to traditional browser cookie based tracking~\cite{RoesnerTracking,Lerner2016InternetJA}, tracking also manifests itself in browser fingerprinting, which relies on browser-specific and OS-specific features and quirks to get a fingerprint of a user's browser that can be used to correlate the user's visits across websites~\cite{Englehardt2016OnlineTA,Cookieless}.
Though tracking can be used for legitimate purposes\dash e.g., preventing fraudulent purchases or providing more personalized content\dash there have been increasing concerns about its privacy implications (e.g., as evidenced by the General Data Protection Regulation, or GDPR, in the EU). At present, though a number of anti-tracking browsers, browser extensions, other tools exist (e.g.,~\cite{privacybadger,ghostery,torbrowser,brave}), there exists no fully reliable way for users to avoid tracking.


\paragraph{Targeting}
Tracking can be and often is ultimately leveraged to target particular content at a user, e.g., via advertisements on Facebook or embedded in other websites (e.g., news sites). Researchers have found that even when advertising platforms like Facebook provide some transparency for users about why certain ads are targeted at them, the information provided is overly broad and vague~\cite{Andreou2018InvestigatingAT}, failing to improve users' understanding of why they are seeing something or what the platform knows about them~\cite{Eslami2018CommunicatingAP}. The act of targeting itself can raise privacy concerns, because it can allow potentially malicious advertisers to extract private information about users stored in advertising platforms (like Facebook)~\cite{Venkatadri2018PrivacyRW,adint,Faizullabhoy2018FacebooksAP}; it can also be used to intentionally or unintentionally target individuals based on sensitive attributes (e.g., health issues)~\cite{Lecuyer2014XRayET}. These very ecosystems, which underlie the economic model of today's web and thus cannot be easily displaced, can be directly exploited by disinformation campaigns to target specific vulnerable users and to create and leverage filter bubbles.



\subsubsection{Social Bots}
\label{sec:bots}

A common tactic seen in recent disinformation campaigns is the use of social media bots to amplify low-credibility content to give the false impression of significant support or interest~\cite{shao2,Ferrara2016TheRO}, or to fan the flames of partisan debate~\cite{StarbirdBLM}. Indeed, bot accounts are common on social media platforms. Some studies estimate that as of 2017, up to 15\% of Twitter accounts are bots~\cite{Varol2017HumanBotI}; Twitter itself recently suspended bot accounts, resulting in a cumulative 6\% drop of followers across the platform~\cite{tw-bots}.

\paragraph{Bot behaviors}
Researchers have characterized the evolution of bots over time, referring to the recent iteration as ``social spambots''~\cite{Ferrara2016TheRO}. Earlier spambots used simplistic techniques\dash such as tweeting spam links at many different accounts~\cite{Cresci2017ThePO}\dash that were relatively easy to detect. By contrast, social spambots use increasingly sophisticated techniques to post context-relevant content based on the community they are attempting to blend into. These bots slowly build trust and credible profiles within online communities~\cite{Ferrara2016TheRO}. They then either programmatically or manually (via a human operator, in a human-bot combination sometimes called a ``cyborg''~\cite{Varol2017HumanBotI}) disseminate false news and extremist perspectives in order to sow distrust and division among human users~\cite{StarbirdBLM}. Social bots have shown to be effective at swaying online discussion as well as inflating the amount of support for various political campaigns and social movements~\cite{Ferrara2016TheRO,shao2}.

\paragraph{Detection and measurement}
Characterizing bot interactions and measuring the social bot ecosystem are critical steps towards being able to detect and eliminate malicious bots. Research into bot ecosystems has found, for example, that unsophisticated bots interact with sophisticated bots to give the sophisticated bots more credibility~\cite{Varol2017HumanBotI}, and that accounts spreading articles from low-credibility sources are more likely to be bots~\cite{shao2}. 

Various efforts have been made to detect and combat social bots online\dash a nice summary is written by Ferrara et al.~\cite{Ferrara2016TheRO}. Detection approaches commonly leverage machine learning to find differences between human users and bot accounts. Detection models make use of the social network graph structure, account data and posting metrics, as well as natural language processing techniques to analyze the text content from profiles. Crowdsourcing techniques have also been attempted~\cite{Wang2013SocialTT}. From the platform provider's vantage point, a promising approach relies on monitoring account behaviour such as time spent viewing posts and number of friend requests sent~\cite{Yang2011UncoveringSN}.

\paragraph{Challenges}
Multiple challenges arise in the fight against malicious social bots. First, some bots are benign or beneficial~\cite{Ferrara2016TheRO,goodbots}, such as bots that help with news aggregation and disaster relief coordination. 
More generally, using machine learning to detect social bots can be challenging due to the lack of ground truth and real world training data, especially since some bots or cyborgs can evade detection even by human evaluators~\cite{Cresci2017ThePO}. Ultimately, bot detection is a classic arms race between attackers and defenders~\cite{Cresci2017ThePO}.

Another challenge is the potential misalignment of incentives: by identifying and banning bot accounts, social media platforms risk depressing their valuations when it becomes clear that the number of legitimate users is lower than previously believed~\cite{fb-bots, tw-bots}. This incentive problem is in tension with the fact that platforms themselves have the best visibility into accounts behaviors that can be effective in detection~\cite{Yang2011UncoveringSN}. Indeed, most academic bot detection work is centered on Twitter, as other platforms are much more difficult to study from an external perspective.
	
Finally, the current mis/disinformation problem is not solely the result of bots; as we have seen, some disinformation campaigns involve sophisticated human operators~\cite{StarbirdBLM}, and the advertising-based economic model of the web encourages clickbait-style content more generally.

\subsection{Technical Defenses}
\label{sec:defenses}

Finally, we summarize existing technical efforts on defending against mis/disinformation, either by attempting to detect it directly (Section~\ref{sec:detection}) and/or by changing the user experience in an information platform (Section~\ref{sec:uiux}).

\subsubsection{Fake News Detection}
\label{sec:detection}

One approach for combating mis/disinformation is to directly tackle the problem as such: detecting fake or misleading news.
Such detection could enable (1)~social platforms to reduce the spread and influence of fake news, and (2)~readers to select reliable sources of information. 

Fake news detection is a space where it is tempting to try to just ``throw some machine learning at the problem'' (e.g.,~\cite{Edell-I-trained,PrezRosas2018AutomaticDO}). However, our exploration of the space suggests that breaking fake news detection down into more carefully defined sub-problems may be more realistic and meaningful at this point.

\paragraph{Definition}
To detect fake news, we first need a formal, well-defined notion of what is ``fake''. A variety of definitions exist in the literature, depending on factors such as whether the facts are verifiably false and whether the publisher's intent is malicious~\cite{Shu2017FakeND}, and including definitions that include or exclude satire, propaganda, and hoaxes~\cite{Rubin2015DeceptionDF}. A commonly used definition~\cite{Shu2017FakeND,fakenewschallenge} of fake news in the context of today's mis/disinformation discussions is information that is both (1)~factually false and (2)~intentionally misleading.

\paragraph{Fake news detection as a unified task}
At first blush, one might approach the problem of fake news detection as a unified task. With machine learning, it is possible to directly train end-to-end models using labeled data (both fake and non-fake articles)~\cite{PrezRosas2018AutomaticDO}. The accessibility of machine learning tools creates a low barrier for attempting this technique~\cite{Edell-I-trained}. However, posing the problem generally in this way comes with pitfalls. First, it requires large amounts of labeled examples which can be costly to obtain; past work has used examples from Politifact~\cite{Wang:17,politifact} and Emergent~\cite{Ferreira:16,emergent} as data sources instead. Additionally, many of these systems are based on neural networks, which lack explanations for their output and therefore make it hard to understand why they do or do not do well, and result in limited usefulness in downstream tasks. Biases in the datasets used can also translate to hidden biases in the classifier, affecting performance when used in real-world situations.

\paragraph{Subtask: Detecting factual inaccuracies}
More effective, particularly from the perspective of advancing the underlying state of the art in natural language processing or related fields, is to solve well-defined sub-problems of fake news detection. One such sub-problem is detecting factual inaccuracies and falsehoods, by examining and verifying the facts in the content of the article based on external evidence. 

A common method deployed in practice is to use experts (fact-checkers) to manually extract the facts and validate each one using any tools necessary (including actively conducting research); examples include Snopes~\cite{snopes} and Politifact~\cite{politifact}. There are also some platforms which use volunteer crowd work to evaluate evidence, for educational purposes (e.g., Mind Over Media~\cite{mindovermedia}) or to maintain veracity of a resource (e.g., Wikipedia).

Using humans to fact-check is, of course, time-consuming and expensive. Automated knowledge-based verification exists but is not yet practical, because it is difficult to formalize claims from natural language, and knowledge bases are incomplete and inflexible with novel information. Recent work has thus focused on even further sub-problems, such as identifying if a claim is backed up by given sources~\cite{Vlachos:14,Derczynski:17,Thorne:18} or using unstructured external web information~\cite{Popat:18}. For example, the Fake News Challenge~\cite{fakenewschallenge,Hanselowski2018ARA} focused on a stance detection sub-problem (based on a prior dataset~\cite{Ferreira:16}) that involves estimating whether a given text agrees, disagrees, discusses, or is unrelated to a headline.


\paragraph{Subtask: Detecting misleading style}
Another sub-problem of identifying fake news is to infer the intent of the article by analyzing its style~\cite{Wang:17}. For example, attempts at deception may be found through identifying manipulative wording or weasel words~\cite{Mihalcea:09,Rashkin:17}, examining the tone of voice (such as identification of clickbait titles), and gauging author inclinations~\cite{Ma:17} and stance~\cite{Ferreira:16,Derczynski:17}. Earlier work on deception detection (e.g.,~\cite{Mihalcea:09}) is also relevant. All of these analyses can be done with either trained human experts or through machine learning models. While writing style offers a strong hint of an author's intent, an absence of manipulative style of writing does not imply good intentions. Conversely, manipulative styles of writing may be associated with other related phenomena, such as hyper-partisanship~\cite{Potthast:18}, rather than fake news directly.

 

\paragraph{Subtask: Analyzing metadata}
Other work has focused on analyzing article metadata rather than article content. For example, when social features, such as shares and likes, are available, they can be used to infer the intent of a collective group. Social features that have been explored include information about the connectivity of the social graph between users, known attributes and profile of the sharer, and the path of propagation of information~\cite{Tambuscio:15,Ma:17,StarbirdEcosystem,Long:17}.


\paragraph{Challenges}
While there is promise in automating fake news detection, there are also possible pitfalls. First, natural (skewed) distribution of fake news can lead to misleading performance numbers of detection systems, as noted in the Fake News Challenge~\cite{fakenewschallenge} postmortem~\cite{Hanselowski2018ARA}. Increasing accessibility of machine learning tools coupled with lack of understanding during evaluation can lead to deployment of systems that are not sufficiently accurate in practice and may have unintended consequences. Secondly, bad actors may be able to leverage fake news detection approaches to create better fake content (as in a generative adversarial network, or GAN), though such an arms race is likely inevitable. 
 
Finally, fake news detection alone, even if done perfectly, does not guarantee a defense against mis/disinformation \dash humans have psychological and cognitive patterns that determine how we actually utilize the information obtained from detection. For example, the backfire effect~\cite{backfireEffect,backfireEffect2} suggests that giving people evidence that something they believe is false can backfire, leading them to believe it more strongly. Thus, an open but important question is: how should the results of fake news detection best be incorporated into tools or interfaces for end users?

\subsubsection{UI/UX Interventions}
\label{sec:uiux}

Researchers and social media platforms have proposed user experience interventions to help users detect online mis/disinformation and reduce the rate at which it spreads. Some of these changes have already been deployed to users, as we describe here. However, it is unclear to the research community whether these interventions have been effective.

\paragraph{Labeling potential mis/disinformation}
One approach is to use the output of fake news detection or similar classifications in order to label low-credibility content. For example, in 2017, Facebook experimented with labeling news articles as ``disputed'' if they were believed false; however this approach was abandoned, as Facebook learned from academics that this does not change people's minds~\cite{fb-disputed}. Instead, Facebook began showing ``related'' articles~\cite{fb-related}, which may include articles from fact-checking sites like Snopes~\cite{snopes}.

\paragraph{More information for users}
Because it is difficult to definitively detect mis/disinformation, and because it can be ineffective to surface this fact to users directly, a common strategy is to simply provide more information to users to help them make their own determination. Facebook's related articles feature~\cite{fb-related} is one example. Other examples from Facebook include more prominently labeling political ads and their funders~\cite{fb-transparent} and providing users with more context about shared articles~\cite{fb-context}. And because hoaxes on Whatsapp have been increasingly shared via forwarded messages, in July 2018, Whatsapp began displaying ``Forwarded'' before such messages to help users understand that the message may not have been written by the immediate sender~\cite{whatsapp-fwd}. 

Third-party tools also exist. For example, the SurfSafe browser extension~\cite{surfsafe,surfsafe-wired} can help users identify doctored images by showing them other websites that the image has been found in \dash so if an image appears newsworthy, but does not appear in any mainstream news sites, users could interpret it as being faked. SurfSafe also highlights whether an image also appears on fact checking sites. The InVID browser extension~\cite{invid} is a set of tools for helping people investigate online content, like stepping frame by frame through videos, reverse image searches, viewing photo/video metadata, and other forensic analysis tools. 

\paragraph{User education}
Another approach Facebook has tried is educating users about how to detect misinformation, and displaying related articles to push users to better educate themselves. In April 2017, Facebook began showing a post titled ``Tips for spotting false news'' the top of users' news feeds, which linked to a page about basic principles of online news literacy, like identifying sources, looking for discrepancies in the URL and formatting, and looking at multiple sources~\cite{fb-edu,fb-tips}. Facebook's related article feature~\cite{fb-related} can also encourage reading articles from multiple sources about controversial topics. Outside of Facebook, researchers have proposed, for example, adding ``nutritional labels'' for news articles to indicate to users which sources are ``healthy'' and ``unhealthy'', like mainstream news versus clickbait~\cite{nutritionLabels}. These labels would display ``health'' metrics like the publication's slant, the tone of the content, the timeliness of the article, and the experience of the author.

\paragraph{Limiting the spread of mis/disinformation}
The designs of different social media platforms may encourage the spread of mis/disinformation, e.g., by rewarding attention-grabbing posts.
Social media platforms have started making changes to their core product to make it less easy to disseminate mis/disinformation. Twitter is considering removing or redesigning the (heart-shaped) ``like'' button, because they are concerned that the incentive to accumulate likes also encourages bad behaviors, which could include spreading sensationalist mis/disinformation~\cite{wired-twitter}. Facebook has made changes to its news feed algorithm to downweight ``clickbait'' posts that abuse sharing mechanisms to reach a larger audience, such as ``share if x'', or ``tag a friend that x''~\cite{fb-clickbait}. This change removes one mechanism that accounts that spread mis/disinformation could use to reach audiences beyond their direct friends and followers.

\paragraph{Challenges}
It is difficult to conduct research on UI/UX interventions, especially for academics, because many critical parts of the research require cooperation from the social media platforms, such as deploying interventions at scale, and measuring the effects of an intervention in the wild. While Facebook and others have made blog posts about the changes they have made, these posts are not scientifically rigorous and detailed enough to be replicable or generalizable. Social media platforms may also be reluctant to work with researchers or experiment with interventions if these conflict or do not align with their business interests. 

Another challenge is that interventions may depend on the accurate identification of mis/disinformation or untrustworthy sources. Doing so automatically remains a significant challenge (as discussed in Sections~\ref{sec:bots} and~\ref{sec:detection}), so defenses like the nutritional labels and SurfSafe still rely on human raters to identify false information.

\section{Discussion}

We now step back from our exploration and summary of the mis/disinformation landscape, and the role of technology in that landscape, to consider first the broader lessons we, as a group, took away from this exploration. We then make recommendations for a variety of stakeholders. 

\subsection{Lessons}
\label{sec:lessons}

\vspace{-0.1in}
\paragraph{Disinformation stems from a variety of motivations and adversarial goals, which may necessitate different solutions} There are many different manifestations of mis/disinformation, and many different actor incentives and intentions~\cite{councilOfEurope}. Thus, different solutions may be required for different actors or manifestations. 
For example, actors with financial incentives may share disinformation to increase click-through profitability, as in the case of Facebook clickbait~\cite{fb-clickbait} or fake news enterprises in Macedonia~\cite{macedonia}. Identifying this motive enables combating it by demonetizing these behaviors (e.g., limiting the spread of clickbait~\cite{fb-clickbait} or erecting barriers in the form of increased manual review~\cite{fb-transparent}). By contrast, state-level actors seeking a political outcome will not be so easily dissuaded; instead, a defensive goal here may be to improve detection and attribution capabilities in order to be able to provide evidence in an international court of law.

\paragraph{Defining the problem is not always straightforward} Given the complexity of this space, the line of what behavior or content is ``bad'' is not necessarily clear or easy to define. Work attempting to combat the issue, e.g., by automatically detecting fake news, should carefully define what the problem is, e.g., what is meant by ``fake news'' \dash and reflect on what these definitions are useful for (and whether they are useful in practice). Moreover, asking platform operators to moderate content is challenging, because in many cases it may be hard to draw a clear line. (There are exceptions, e.g., human rights atrocities~\cite{mulligan}.) 

\paragraph{Solutions can and will not just be technical} The motivations, forms, and methods for spreading mis/disinformation are diverse, complicated and evolving;  automated approaches for detecting it seem insufficient.
Moreoever, mis/disinformation exploits underlying human congitive vulnerabilities. Thus, we cannot just ``throw some machine learning'' at the problem but must consider more holistic approaches. That said, some promising technical approaches do exist and should be further researched and implemented. We make some recommendations in the next section.

\paragraph{The disinformation problem is an arms race, and an asymmetrical battle between attackers and defenders} Mis/disinformation is a difficult problem to fight because it is a highly asymmetrical battle. Like insurgents, spreaders of online disinformation have a wide variety of established targets to attack and the freedom to appear, test new strategies, and disappear again with minimal cost and few repercussions. Meanwhile, defenders like fact-checkers have a much more difficult task ahead of them, that is both hard to define and often contrary to people's cognitive biases. In such a complicated space, it may be unsurprising that we cannot indiscriminately point technology at the problem and make it go away; what is worrying, though, is that it seems to be very easy for adversaries to indiscriminately use technology and make it much worse.

\paragraph{Current adversaries are already sophisticated, and can/will become more so; but they also use simple yet effective techniques} A number of case studies show how sophisticated disinformation campaigns already are in some cases, e.g., creating targeted content for each group of users that resonates with their opinions~\cite{StarbirdBLM}.  Disinformation is becoming more realistic, more personalized, and more widespread \dash including convincing fake videos and social media personas. At the same time, sometimes the most effective disinformation is not at all sophisticated in terms of technology. For example, complex deep learning video editing techniques are not required for simple edited photos that effectively spread virally (e.g., a fake photo of an NFL player burning an American flag~\cite{NFL}). However, like spear phishing~\cite{spearphishing} in the computer security domain, such simple techniques must be taken seriously. Additionally, the simplicity of some of the known techniques \dash like copying article text~\cite{StarbirdEcosystem} \dash contributes to the impression that this is an arms race, and detecting disinformation now will not mean we can detect disinformation tomorrow.

\paragraph{Incentives of platforms, and even users themselves, are often not aligned with combating mis/disinformation} It is difficult to credibly make recommendations to platform designers that are employed by for-profit companies: these companies (and by extension, the designers) may prioritize profit if it conflicts with combating mis/disinformation. This concern is not just academic: Facebook's and Twitter's user numbers, for example, may directly impact their valuations~\cite{fb-bots,tw-bots}, reducing the incentive to separate legitimate from illegitimate users.
Though these platforms are not designed to intentionally support mis/disinformation, often design choices interact badly with human cognitive biases. We discuss this issue further in our recommendations below.

\paragraph{Solutions should be careful about the risk of causing people to be overly skeptical and distrust everything} Sowing distrust and undermining institutions is a key goal of some disinformation campaigns, and solutions must be careful not to ``throw the baby out with the bathwater'' \dash that is, users must be taught to and supported in building (legitimate) trust, not just skepticism.

\paragraph{There may be lessons to apply from traditional computer security} For example, lessons from usable security suggest that interrupting users and asking them to make decisions about what is good or bad is not, in and of itself, an effective strategy~\cite{usec}. As another example, the best technical systems can be undermined by targeted spear phishing attacks~\cite{spearphishing}, targeting human cognitive vulnerabilities.
\subsection{Recommendations}
\label{sec:recommendations}

Finally, we surface our recommendations, organized by stakeholder.

\subsubsection{Recommendations for Technologists and Other Defenders}

\vspace{-0.1in}
\paragraph{Be specific about the goals that a particular technical solution is intending to achieve} If one's goal is to solve a problem, then one should focus on the goal, not the technology used to solve it. A good goal is both useful and possible to achieve. 
A useful example is the work of Starbird et al.~\cite{StarbirdEcosystem}, which identified a useful task (investigating the ecosystem) and published valuable data and insights as a result. Another possibly instructive comparison is between the Fake News Challenge~\cite{Hanselowski2018ARA}, which pushed forward the state of the art on a specific problem (stance detection) and other, more general attempts at fake news detection with machine learning~\cite{PrezRosas2018AutomaticDO,Edell-I-trained}, whose general applicability and ultimately effectiveness may be limited. 

\paragraph{Do work on solving technology-related sub-problems}
Though technology alone cannot solve the whole problem, as discussed, there are technology-related sub-problems that are useful.
Example technology-based directions to investigate (though not an exhaustive list) may include:
\begin{itemize}
\item False news detection based on precise definitions of useful, solvable sub-problems; combining these sub-problems into a larger detection pipeline; and leveraging the results of this detection (e.g., to measure ecosystems or effectively surface information to end users).
\item ``Chain of custody'' approaches for information. For example, for photos, see ProofMode~\cite{proofmode}, TruePic~\cite{truepic}, and PhotoProof~\cite{photoproof}. 
\item Platform-based or other UI/UX interventions: both formative research to understand what is (not) effective and how existing designs drive user behaviors, and developing and evaluating new designs.
\item Continued work on bot detection and characterization, and on strategies for user verification.
\end{itemize}

\paragraph{Continue measurement studies to understand the ecosystem} Successful defenses require a deep understanding of the underlying ecosystems creating, spreading, and reacting to mis/disinformation.

\paragraph{Tailor solutions to specific types of users, content, and/or adversaries} Different parts of the space may necessitate different solutions, as discussed above.

\paragraph{Consider what alternate incentive or monetary structures for the web might be, and how to make those a reality} The financial incentives driving platform providers and many adversaries are also core to the functioning of the modern web, e.g., the supporting of “free” content and services via targeted advertisements. Though challenging, we must consider alternate models that will de-incentivize financially-based mis/disinformation and better align the incentives of platform providers, users, content creators, and mis/disinformation defenders.

\paragraph{Invest in and study user education efforts} Skills like critical thinking and the ability to fact check are good defenses against mis/disinformation in general, but applying them becomes more challenging as technology enables more sophisticated and subtle mis/disinformation. We should explore how to effectively teach users how current technologies could be exploited, and how to spot these issues. For example, this could include: 
\begin{itemize}
\item Teaching users to spot ads versus organic content.
\item Teaching users to be aware of what technology is capable of (e.g., screenshots can be easily fabricated, fake portraits and videos made, accounts hacked). 
\item Teaching users to look for markers of bot accounts and low-credibility stories and sources. 
\item Teaching users to spot mis/disinformation via a game.
\item Balancing teaching skepticism with teaching how to build up knowledge. The end goal should not be for users to simply distrust everything.
\item Bringing discussions of this topic and possible interventions into the ``public square'', such as via television shows. There may be models for this type of engagement and/or lessons to learn from other attempts, such as the Latvian ``Theory of Lies'' show on uncovering disinformation~\cite{theoryoflies} or the bilingual ``StopFake'' show in the Ukraine~\cite{stopfake}.  

\end{itemize}
We also recommend following a scientific process to develop and evaluate educational interventions. Researchers and others should test and evaluate which specific educational interventions actually have a positive effect.

\paragraph{Avoid placing too much burden on users} 
Though education and surfacing more information to users can be a part of the solution, it can certainly not be the entire solution. Taking lessons from, for example, the usable computer security community (e.g.,~\cite{usec}), solutions that increase the burden on the user and get in the way of a user's primary task (in this case, interacting with their social network and consuming information) may be circumvented, ignored, and/or ultimately ineffective. Thus, user education must be complemented by other interventions discussed elsewhere in this section.

\paragraph{Consider all users} We need to develop solutions for everyone and not only the technically literate 1\%. Fact-checking tools and UI interventions only work if people actually use them. Thus, for example, browser extensions or other solutions that require users to go out of their way will have limited reach. As a result, some burden may be on social media platform designers, news sites and aggregators, and browser vendors themselves to consider what UI/UX interventions may benefit all users. We discuss specific suggestions for (particularly social media) platform designers in the next section. Additionally, different geographic or demographic user groups may interact differently with mis/disinformation and interventions.

\subsubsection{Recommendations for Platform Designers}

\vspace{-0.1in}
\paragraph{Acknowledge that platform designs are not neutral} 
Designers should recognize that some design patterns (while useful for user engagement) could have unintended consequences and leave openings for mis/disinformation. Most prominently, patterns for generating revenue (tracking, ad targeting, and so on) can and are misused by malicious parties. As another example, Twitter's emphasis on a username instead of the unique ``@handle'' enables attacks in which compromised ``verified'' accounts can be used to masquerade as other prominent users~\cite{fake-elon}.  
Platform designers are in a unique position to directly influence the mis/disinformation landscape.
Some concrete recommendations for changes to investigate:
\begin{itemize} 

\item Platforms should examine their design affordances and study their role in mis/disinformation. For example, consider some of the social signals that their platforms introduce, such as likes, shares, and the selection of comments shown. In their current form, while they show what is popular, these UI features also confer credibility and encourage the spread of falsehoods and propaganda. Some platforms have already begun to make changes to these features, though their impacts are not publicly known.

\item Platforms should be more thoughtful about their recommender algorithms, and continue to find ways to downweight content that is extreme and misleading. 

\item Platforms should continue to provide users with more information and more (genuine) transparency about, for example, the sources of information, the reasons for why certain content is targeted at a user, and which content is sponsored versus organic. Prior works suggests that existing efforts to disclose and label advertisements is ineffective (e.g.,~\cite{Mathur2018EndorsementsOS}).

\item Consider how identity is handled on the platform. 
Though there is value in places on the web where things can be shared anonymously, its use\dash and how illegitimate accounts are identified and handled\dash should be considered carefully. As a cautionary tale, consider how easily Russian actors impersonated members of the Black Lives Matter movement~\cite{StarbirdBLM}. 
One possible mitigation to study is a design where posts from untrusted, anonymous sources are not intermingled with content authored directly by accounts known to a user. 

\item More generally, platform designers should try to shift the incentives of their organizations to treat combating mis/disinformation as a higher high priority than some reports suggest it is today (e.g.,~\cite{fb-snopes}), recognizing that this phenomenon, if unaddressed, could ultimately pose an existential threats to these platforms. 

\end{itemize}

\paragraph{Provide ways for users/researchers/others to understand and evaluate what is going on in the platform} One possibility may be teaming up with researchers to experiment with and understand the effects of each feature, both good and bad. Some such efforts already exist, such as Facebook providing data to researchers via SocialScienceOne~\cite{socialscienceone}.

\subsubsection{Recommendations for Researchers}

\vspace{-0.1in}
\paragraph{Make the results of research in this space accessible to and digestible by the general public} We know there is a problem. Research, such as that summarized in Section~\ref{sec:measure}, clearly documents the problem. And the results of this research needs to continue to move out of academia to the general public, to support broader discussion and awareness. 

\paragraph{Think critically about the technologies we develop and how they might be misused} While technologies help enable mis/disinformation, most of them are designed with certain other motivations. For instance, tracking and targeting is primarily for personalization, computer-generated video improves the quality of post-production, and bots can also be benign. We, as researchers, must acknowledge that technological advances have also enabled the creation and spread of mis/disinformation at a massive scale. When developing a system or technology, we must assume bad actors will attempt to exploit or otherwise take advantage of these systems; pretending that people will not misuse technologies, e.g., manipulate videos for propaganda purposes, is naive. 

\paragraph{Collaborate across fields to understand and address the problem} It is critical that researchers from multiple (sub)fields, including 
cognitive science,
computer security, 
behavioral sciences,
game theory,
human-computer interaction,
information science, 
journalism,
law,
machine learning,
natural language processing,
neurology,
psychology,
sociology,
and others work together to understand and address this problem. Otherwise proposed solutions risk being ineffective due to an incomplete understanding of the issues. For example, formulating defenses against mis/disinformation as a supervised binary classification problem may have major limitations in practice \dash e.g., due to the backfire effect~\cite{backfireEffect,backfireEffect}, flagging content as fake can make users more likely to believe it, not less. Instead, detection, prevention, and defense against mis/disinformation is an interdisciplinary research problem.

\paragraph{Study different groups of users.} Researchers should work to understand how the effect of mis/disinformation and effective offensive/defensive strategies may differ between different users groups. As with different adversarial motivations, different approaches may be needed for different users. Foundational science is still needed to understand these potential differences and their impacts.

\subsubsection{Recommendations for Policymakers}

\vspace{-0.1in}
\paragraph{Consider policies or regulations that might help align the incentives of platform developers to combating the mis/disinformation problem}
Making recommendations to policymakers was not the goal of our investigation, so we do not intend to present a complete list here. However, as we often came up against the tensions between the web's economic model (via advertisements) and the viability of possible platform-based interventions to limit the spread of mis/disinformation, we felt this point was important to mention. We suggest that policymakers should consider their potential role in helping reshape the incentive structures that lead to this tension\dash for example, in regulating how users' data can be used for targeting or how platforms should respond to known disinformation campaigns. 

\paragraph{Ground policy proposals in technical realities}
At the same time, we caution that policymakers should ground their proposals in a solid understanding of the underlying technologies. For example, a proposed U.S. Senate bill intending to regulate bots by requiring them to self-identify (and platform providers to enforce it)~\cite{bot-act} suffers from an imprecise definition of bot~\cite{bots-slate} and may be technically challenging or impossible to enforce, sweep up legitimate users or use cases, fail to positively affect end user behaviors even if enforced, and may thus ultimately burden platforms and users without having the intended benefits.

\subsubsection{Recommendations for Users}

Additional and more comprehensive recommendations for users can be found online (e.g.,~\cite{users1,users2,users3,fb-tips}). We focus briefly on several high-level recommendations.

\paragraph{Be skeptical of content you see on the web} Content from unknown sources, or that seems suspicious or particularly dramatic or too good or bad to be true, should be taken with a healthy dose of skepticism. Even further consideration should be taken before sharing such content with others. Be aware that even ``trusted'' sources of information can reshare mis/disinformation; your friends may inadvertently do so, and not all platforms surface information about the original source (e.g., forwarded messages on WhatsApp). 
Look for corroborating or conflicting evidence from other sources, and be open to changing your mind.

\paragraph{At the same time, don't assume everything is false just because it \textit{can} be false} Undermining the legitimacy of institutions like journalism is a key goal of some disinformation campaigns. Use strategies to build your trust in content: for example, try to identify the original sources, go directly to news sources you trust, and seek out a diversity of sources. 

\paragraph{Be aware of your own possible cognitive biases and try to spot when they are being exploited} For example, be cautious if something makes you angry and confirms your existing view about a person or issue. An instructive list of cognitive biases can be found in the references~\cite{cognitivebiases}.
\section{Conclusion}

The current ecosystem of technology-enabled mis- and disinformation has potentially serious and far-reaching consequences. As technologists, it is critical that we understand the potential and actual negative impacts and misuses of the technologies we create; that we deploy defensive technologies at aimed at specific, achievable goals; and that we collaborate with other fields to fully understand the problem and solution space. We have compiled this report, the result of an academic quarter's investigation into the mis/disinformation landscape from the perspective of a graduate special topics course in computer science, in the hopes that the interested reader will benefit from our summaries, lessons, and recommendations. Further efforts from researchers, technologists, platform designers, educators, policymakers, and others are required to study and combat both intentional disinformation and unintentional misinformation. Ultimately this is not a problem that can be eliminated completely\dash the underlying human cognitive biases and their exploitation for persuasion, both good and bad, long predate the technologies discussed in this report. But that does not mean that we should throw up our hands and avoid thinking critically about how to improve the information ecosystems that we use and build.

{\small \bibliographystyle{acm}
\bibliography{bib}}

\end{document}